\documentclass[12pt]{article}
\usepackage{amsfonts}
\begin{document}
\begin{center}
\large{\bf{On the Algebraic Bethe Ansatz for XXX spin chain:\\ creation
operators "beyond the equator".}}

\vspace*{10mm}

{\bf A.E. Kovalsky}\\
{\it MIPT, Dolgoprudny, Moscow reg., \\
 IHEP, Protvino, Moscow reg., Russia,} \\
{\bf G.P. Pronko}\\
{\it IHEP, Protvino, Moscow reg., Russia,\\
International Solvay Institute, Brussels, Belgium}
\end{center}

\begin{abstract}
Considering the XXX spin-$1/2$ chain in the framework of the Algebraic
Bethe
Ansatz (ABA) we make the following short comment:  the product of the
creation
operators corresponding to the recently found solution of the Bethe
equations
"on the wrong side of the equator" \cite{PrStr} is just zero (not only its
action on the pseudovacuum).
\end{abstract}

\vspace*{10mm}

 Consider the periodic XXX spin-$1/2$ chain with $N$ sites in the
 framework of
 ABA (see for example \cite{Faddeev}).
 Let us introduce the Lax operator acting in the two-dimensional local
 quantum
 space $h_n=\mathbb C^2$ and in the two-dimensional auxiliary space
 $V=\mathbb C^2$:
 \begin{eqnarray}
 L_n(x)=\left(\begin{array}{cc} x+i s_n^3 & i s^-_n \\ i s^+_n  &  x-i
 s^3_n 
 \end{array}\right),
\label{Lax}
\end{eqnarray}
 where $s^{i}$ are operators of spin $1/2$, $x$ is an arbituary complex
 number
 (the spectral parameter).
 The monodromy matrix is the ordered product over all sites:
\begin{equation}
T(x)=L_N(x) L_{N-1}(x)...L_{1}(x)=\left(\begin{array}{cc} A(x) & B(x)
\\C(x) &
D(x) 
 \end{array}\right),
 \end{equation}
 where $ A(x), B(x), C(x), D(x) $ are operators acting in the full quantum
 space $H=\otimes \prod\limits_{n=1}^{N} h_n$. In the framework of ABA one
 looks for the eigenvectors of the transfer matrix $$\hat t(x)=tr
 T(x)=A(x)+D(x)$$  in the form
 \begin{equation}
 \Phi(\{x\})=B(x_1)B(x_2)...B(x_l) \Omega,
 \label{bethevector}
 \end{equation}
where $\Omega=\prod\limits_{n=1}^{N}\omega_n,~s^+_n \omega_n=0$.
It follows from the intertwining  relations for the monodromy matricies
that
vector (\ref{bethevector}) will be an eigenvector of the transfer matrix
when
the parameters $x_1,...,x_l$ satisfy the Bethe equations:
\begin{equation}
\left(\frac{x_j+i/2}{x_j-i/2}\right)^N=\prod\limits_{k\neq
j}^{l}\frac{x_j-x_k+i}{x_j-x_k-i}~~~~(j=1,2,...,l).
\label{system}
\end{equation}

Let us denote vectors of the form (\ref{bethevector}) with parameters
$x_j~(j=1,2,...,l)$ satisfying the system (\ref{system}) by
$\Phi(\{x\}_{B})$.
It is well known that vectors $\Phi(\{x\}_{B})$ are the highest weights
vectors
with respect to $SU(2)$ generated by $J^i$ i.e. 
\begin{equation}
J^+ \Phi(\{x\}_{B})=0
\end{equation}
and 
\begin{equation}
J^3 \Phi(\{x\}_{B})=(N/2-l)\Phi(\{x\}_{B}),
\end{equation}
 where $J^{3},~J^{\pm}$ are operators of the total spin. It is clear that
 if
 $l>N/2$ then  $\Phi(\{x\}_{B})=0$. 
 The solution $\{x\}$ of (\ref{system})  with $l \leq N/2$ defines the
 polynomial $q(x)$ of the degree $l$, whose roots are $\{x\}$:
 \begin{equation}
 q(x)=\prod\limits_{j=1}^{l} (x-x_j).
 \end{equation}

 Let $t(x)$ be the eigenvalue of transfer matrix $\hat t(x)$ corresponding
 to
 the eigenvector $\Phi(\{x\}_{B})$ i.e. $\hat t(x) \Phi(\{x\}_{B})=t(x)
 \Phi(\{x\}_{B})$. It is a polynomial of degree $N$.
 Then the polynomials $t(x)$ and $q(x)$ satisfy the Baxter equation
 \cite{Baxter} (we consider the case of the simple roots):
 \begin{equation}
  t(x) q(x)=(x-i/2)^N q(x+i)+(x+i/2)^N q(x-i).
 \label{Baxter}
 \end{equation}
 
 In the paper \cite{PrStr} was shown that there exist the polynomial
 $p(x)$ of
 degree $N-l+1$ with roots satisfying the Bethe equation (\ref{system})
 \footnote{See also the interesting discussion of the "beyond equator"
 solution
 in \cite{B2}.} and 
 \begin{equation}
  t(x) p(x)=(x-i/2)^N p(x+i)+(x+i/2)^N p(x-i),
 \label{Baxter2}
 \end{equation}
 with the same $t(x)$ as in (\ref{Baxter}). Actually there exists the
 one-parametric family of such polynomials 
 \begin{equation}
 p(x,\alpha)=p(x)+\alpha q(x),
 \label{dep}
 \end{equation}
 so there is the one-parametric family of sets of parameters
 $(\{x\})$--the
 zeroes of $p(x,\alpha)$ which belongs to the "beyond the equator" case.
 Let us
 denote these zeroes as $x_i(\alpha)$ (it is clear that the zeroes of the
 polynomial (\ref{dep}) depends on $\alpha$). Now consider the creation
 operator 
 $${\bf B }(\alpha)= B(x_1(\alpha))
 B(x_2(\alpha))...B(x_{N-l+1}(\alpha))$$
 corresponding to the beyond the equator case. The following statement is
 valid:\\
 {\bf Theorem.}
  \begin{equation}
 {\bf B }(\alpha)=0
 \end{equation}
  {\bf Proof.}\\
   Consider its action on the basis constructed using the Bethe vectors
   (in the
   case of finite $\{x\}$ Bethe vectors are the highest weights; to obtain
   the
   rest eigenvectors  we use the operator $J^{-}$ (it commutes with the
   transfer matrix), which also can be considered as a creation operator,
   since
   $B(x)=x^{N-1} (J^{-}+o(1/x) )$ when $x\to\infty$; on the hypothesis of
   the completeness of the Bethe ansatz see \cite{Faddeev, Kir}). We
   immidiatly
   see that
   its action is zero due to $[B(x),B(y)]=0,~[B(x),J^{-}]=0$ and ${\bf B
   }(\alpha) \Omega=0$. So, the action of ${\bf B }(\alpha)$ is zero onto
   each
   vector of the basis, then 
 \begin{equation}
 {\bf B }(\alpha)=0.
 \label{result}
 \end{equation}
 
  To see that this fact is nontrivial let us consider the concrete
  examples.
  Let us first analyse the structures of the product of $B$-operators. We
  have
  the following commutation relation:
  \begin{equation}
  [B(x),J^3]=B(x),
  \end{equation}
   so for the product of $l$ $B$-operators
  \begin{equation}
  e^{\beta J^3} B(x_1)...B(x_l) e^{-\beta J^3}=e^{-l \beta}
  B(x_1)...B(x_l)
 \label{restriction} 
 \end{equation}
  and we see that each term of expansion of this product necessarily
  contains
  the products of  $l$-operators $s^-_j $ with different $j$, since
  $(s^-)^2=0$
  (so  if $l>N$ this product is zero).  At $N=6$, product of four
  $B$-operators
  ($l=4>3$ i.e. this is the "beyond the equator" case): $B(x_1) B(x_2)
  B(x_3)
  B(x_4)$, as was shown by the explicit construction of this product,
  contains
  terms, proportional to $s^+_2 s^-_1 s^-_3 s^-_4 s^-_5 s^-_6$, $s^+_3
  s^-_1
  s^-_2 s^-_4 s^-_5 s^-_6$,  $s^+_4 s^-_1 s^-_2 s^-_3 s^-_5 s^-_6$ and
  $s^+_5
  s^-_1 s^-_2 s^-_3 s^-_4 s^-_6$ with  nonzero coefficients--polynomials
  in
  $x_1,x_2,x_3,x_4$. The appearance of such terms is not excluded by
  (\ref{restriction}). Its action on the vacuum $\Omega$ is zero
  identicaly, so
  if $x_1,x_2,x_3,x_4$ satisfy the system (\ref{system}), then these terms
  is
  zero, not only terms, whose action on the vacuum $\Omega$ is nonzero,
  for
  example the term proportional to
  $s^-_1 s^-_2 s^-_3 s^-_4$. Such solutions do exist, for example  roots
  of the
  polynomial 
  \begin{equation}
  p(x)=x^4 - \frac{6}{ \sqrt{13}}~ x^3+ x^2 -\frac{9}{16}
  \label{example}
  \end{equation}
  satisfy system (\ref{system}).
  This solution corresponds to the total spin $J=0$, the eigenvalue of the
  transfer matrix
  $$t(x)=2 x^6+ \frac{9}{2}~
  x^4+\frac{23}{8}~x^2-\frac{3}{\sqrt{13}}~x-\frac{1}{32}$$
  the corresponding eigenvector can be constructed, using the roots of the
  polynomial
  \begin{equation}
  q(x)=x^3+\frac{1}{12}~x+\frac{1}{4 \sqrt{13}}~.
  \label{example2}
  \end{equation} 
  and the one-parametric family ${\bf B }(\alpha)=0$ corresponds to the
  roots
  of the polynomial 
  \begin{equation}
  p(x,\alpha)=x^4 - \frac{6}{ \sqrt{13}}~ x^3+ x^2
  -\frac{9}{16}+\alpha(
  x^3+\frac{1}{12}~x+\frac{1}{4 \sqrt{13}}).
  \end{equation}
  
  The products of $l$ operators $B(x)$ with $l > N/2$ considered above are
  not used for the construction of the eigenvectors of the transfer
  matrix.
  However, we would like to emphasize that there is another important
  example,
  when the product of $B$-operators corresponding to the case $l \leq N/2$
  is
  zero:
  $$B(-i/2)B(i/2)=0.$$ 
  This product corresponds to the polynomial $$q(x)=x^2+1/4$$ and
  $$t(x)=(x+i/2)(x-3/2 i)^{N-1}+(x-i/2)(x+3/2 i)^{N-1}$$ and at even $N
  \geq 4$
  corresponds to some eigenvector of the transfer matrix. In the paper
  \cite{AvVlad} was shown how one can use the ABA to construct this
  eigenvector
  corresponding to this exceptional solution. If we consider the following
  vector 
  $B(-i/2+\epsilon+2 i \epsilon^N)B(i/2+\epsilon) \Omega$ at $\epsilon \to
  0$,
  then 
  $$
  B(-i/2+\epsilon+2 i \epsilon^N)B(i/2+\epsilon) \Omega =\epsilon^N
  \Phi(\{-i/2,i/2\})+O(\epsilon^{N+1})
  $$
  where  $\Phi(\{-i/2,i/2\})$ is the desired eigenvector. The proof of our
  statement holds true in this case too, since we use again only
  $B$-operators
  to construct this eigenvector.
 \vspace*{5mm}

  This work was supported in part by grants of RFBR 00-15-96645,
  01-02-16585, 01-01-00201, CRDF MO-011-0, of the Russian Minestry on the
  education
  E00-3.3-62 and INTAS 00-00561.


\begin{thebibliography}{**}

\bibitem{PrStr}
Pronko G.P., Stroganov Yu.G.,\\ 
Bethe Equations "on the Wrong Side of Equator",	 \\
J.Phys.A: Math.Gen. {\bf 32} 2333-40, 1999. \\
e-Print Archive: hep-th/9808153. 

\bibitem{Faddeev} 
Faddeev L. D., \\
How Algebraic Bethe Ansatz works for integrable model.\\
1995 $UMANA$ {\bf{40}} 214,\\ e-Print Archive: hep-th/9605187.



\bibitem{Kir}
A.N. Kirillov, \\
Combinatorial identities and completeness of states for \\
the generalized Heisenberg magnet,\\ 
Zap. Nauch. Sem. LOMI, 1984, v.131, p.88-105.

\bibitem{Baxter}
Baxter R.J. Stud.Appl.Math, L51-69, 1971;
     Ann.Phys. N.Y., v.70, 193-228, 1972;
     Ann.Phys. N.Y., v.76, 1-71, 1973.

\bibitem{B2}
R.J. Baxter, \\
Completeness of the Bethe ansatz for the six and eight-vertex models\\
e-Print Archive: cond-mat/0111188 

\bibitem{AvVlad}
L. V. Avdeev, A. A. Vladimirov, \\
On the exceptional solutions of the Bethe-Ansatz equations.\\
Theor. Math. Phys. {\bf 69}: 1071, 1987. 



\end{thebibliography}
\end{document}